# The AGEL Survey: Spectroscopic Confirmation of Strong Gravitational Lenses in the DES and DECaLS Fields Selected Using Convolutional Neural Networks

Kim-Vy H. Tran[1,2], Anishya Harshan[1,2], Karl Glazebrook[2,3], G. C. Keerthi Vasan[4], Tucker Jones[4], Colin Jacobs[3], Glenn G. Kacprzak[2,3], Tania M. Barone[2,3], Thomas E. Collett[5], Anshu Gupta[6,10], Astrid Henderson[1], Lisa J. Kewley[2,7], Sebastian Lopez[8], Themiya Nanayakkara[2,3], Ryan L. Sanders[4], and Sarah M. Sweet[2,9]

[1] School of Physics, University of New South Wales, Kensington, Australia; kimvy.tran@gmail.com
[2] ARC Centre for Excellence in All-Sky Astrophysics in 3D (ASTRO 3D), Australia
[3] Swinburne University of Technology, Hawthorn, VIC 3122, Australia
[4] Department of Physics and Astronomy, University of California, Davis, One Shields Ave, Davis, CA 95616, USA
[5] Institute of Cosmology and Gravitation, University of Portsmouth, Burnaby Rd, Portsmouth PO1 3FX, UK
[6] International Centre for Radio Astronomy Research (ICRAR), Curtin University, Bentley, WA, Australia
[7] Research School for Astronomy & Astrophysics, Australian National University, Canberra, ACT 2611, Australia
[8] Departamento de Astronomía, Universidad de Chile, Casilla 36-D, Santiago, Chile
[9] School of Mathematics and Physics, University of Queensland, Brisbane, QLD 4072, Australia
*Received 2021 December 19; revised 2022 June 27; accepted 2022 June 29; published 2022 September 26*

## Abstract

We present spectroscopic confirmation of candidate strong gravitational lenses using the Keck Observatory and Very Large Telescope as part of our ASTRO 3D Galaxy Evolution with Lenses (AGEL) survey. We confirm that (1) search methods using convolutional neural networks (CNNs) with visual inspection successfully identify strong gravitational lenses and (2) the lenses are at higher redshifts relative to existing surveys due to the combination of deeper and higher-resolution imaging from DECam and spectroscopy spanning optical to near-infrared wavelengths. We measure 104 redshifts in 77 systems selected from a catalog in the DES and DECaLS imaging fields ($r \leqslant 22$ mag). Combining our results with published redshifts, we present redshifts for 68 lenses and establish that CNN-based searches are highly effective for use in future imaging surveys with a success rate of at least 88% (defined as 68/77). We report 53 strong lenses with spectroscopic redshifts for both the deflector and source ($z_{\rm src} > z_{\rm defl}$), and 15 lenses with a spectroscopic redshift for either the deflector ($z_{\rm defl} > 0.21$) or source ($z_{\rm src} \geqslant 1.34$). For the 68 lenses, the deflectors and sources have average redshifts and standard deviations of $0.58 \pm 0.14$ and $1.92 \pm 0.59$ respectively, and corresponding redshift ranges of $z_{\rm defl} = 0.21$–0.89 and $z_{\rm src} = 0.88$–3.55. The AGEL systems include 41 deflectors at $z_{\rm defl} \geqslant 0.5$ that are ideal for follow-up studies to track how mass density profiles evolve with redshift. Our goal with AGEL is to spectroscopically confirm ∼100 strong gravitational lenses that can be observed from both hemispheres throughout the year. The AGEL survey is a resource for refining automated all-sky searches and addressing a range of questions in astrophysics and cosmology.

*Unified Astronomy Thesaurus concepts:* Strong gravitational lensing (1643); Galaxy evolution (594); Spectroscopy (1558); Redshift surveys (1378); Galaxy formation (595); Optical astronomy (1776)

## 1. Introduction

Gravitational lenses are powerful cosmic magnifying glasses that we now regularly use to explore a wide range of astrophysical phenomena. Strong gravitational lensing extends our observational reach to include objects that are too faint for even the most powerful telescopes with the added bonus of spatially resolving internal structures of distant objects at subkiloparsec scales. By tracing the total matter distribution, gravitational lensing also illuminates dark matter halos of foreground deflectors that span the range from single galaxies to galaxy clusters up to $z = 1.62$ (Franx et al. 1997; Sonnenfeld et al. 2013; Wong et al. 2014). With high-resolution observations from the Hubble Space Telescope (HST) at optical/near-IR wavelengths and the Atacama Large Millimeter/submillimeter Array at longer wavelengths, strong gravitational lensing has enabled multiple imaging of a single supernova, discovery and analysis of galaxies with the highest redshift, mapping of dark matter distributions from the subkiloparsec to megaparsec regime, and measurement of the Hubble constant (e.g., Jones et al. 2013; Kelly et al. 2015; Yuan et al. 2015; Leethochawalit et al. 2016; Oesch et al. 2016; Meneghetti et al. 2017; Suyu et al. 2017).

Identifying strong gravitational lenses has been challenging due to the required combination of high-resolution imaging, wide-area surveys, and spectroscopic confirmation (Bolton et al. 2008; Gavazzi et al. 2012; Stark et al. 2013). Lenses have complex morphologies, and flux from the foreground deflector and background source is usually blended in ground-based observations. Subarcsecond imaging is key to detecting the distinctive visual signature of gravitational arcs and rings, and spectroscopic follow-up is needed to confirm the foreground lens and background source. Bright ($r \lesssim 22$ mag) gravitational lenses that can be followed up with adaptive optics are ideal for multiwavelength observations at high spatial or spectral resolution. However, bright lenses are rare ($\lesssim 0.1$ per square degree; Jacobs et al. 2019a, 2019b; Huang et al. 2020), and identifying more than a handful requires imaging hundreds of square degrees (see also SL2S; Gavazzi et al. 2012).

---
[10] ARC Centre for Excellence in All-Sky Astrophysics in 3D (ASTRO 3D) Fellow.

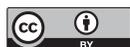

The Sloan Digital Sky Survey (SDSS) made possible the first generation of wide-area searches for strong gravitational lenses, but the galaxy-scale lenses studied thus far are not representative of the broader population. Most lensing candidates in SDSS were identified using fiber spectroscopy that captured light from both the deflector and source (e.g., SLACS and BELLS; Bolton et al. 2008; Brownstein et al. 2012), and thus are limited to lenses with Einstein radii ($r_{EIN}$) of $\lesssim 1\rlap{.}''5$ due to the fiber diameter of $3''$. Fiber searches miss wide single galaxy lenses like the Cosmic Horseshoe ($r_{EIN} = 5''$, $z_{lens} = 0.44$; Belokurov et al. 2007) and group/cluster-scale lenses. SDSS-based searches also have a magnitude limit of $i < 20$ mag, which means that most of the confirmed galaxy-scale (foreground) deflectors are at $z \lesssim 0.6$ (Bolton et al. 2008; Brownstein et al. 2012; Stark et al. 2013). Complementary searches targeting larger lenses ($r_{EIN} > 3''$) in SDSS such as CASSOWARY (Belokurov et al. 2009; Stark et al. 2013) and RCS (Bayliss et al. 2011) are also limited to the SDSS depth and resolution.

Here we introduce our ASTRO 3D Galaxy Evolution with Lenses (AGEL) survey to spectroscopically confirm strong gravitational lenses selected from deep optical imaging with the Dark Energy Survey (DES; Abbott et al. 2018) and Dark Energy Camera Legacy Survey (DECaLS; Dey et al. 2019) using convolutional neural networks (CNNs; Jacobs et al. 2019a, 2019b, hereafter jointly J19ab). The DECam imaging available in these public surveys reaches fainter magnitudes and has better angular resolution than SDSS, qualities thereby enabling the AGEL survey to push to volumes at higher redshift and detect gravitational lenses with $r_{EIN} > 1\rlap{.}''5$. CNN-based methods can efficiently sift through increasingly larger data sets like DES to search for the distinct visual signature of gravitational lensing, a process that would be virtually impossible with the human eye alone (Metcalf et al. 2019). We build on earlier searches that used human inspection (More et al. 2016; Diehl et al. 2017), lens modeling (Chan et al. 2015), or neural networks (Jacobs et al. 2017; Petrillo et al. 2017; Huang et al. 2020) to identify high-quality gravitational lenses and increase the number of candidates from the hundreds to the thousands.

Developing neural networks (NNs) to produce high-fidelity and high-purity catalogs for different classes of objects is important because upcoming deep, wide-field surveys such as EUCLID and LSST will discover $> 10^4$ lensing systems (Metcalf et al. 2019). NNs are critical for sifting through millions of objects to identify a few thousand candidates that can then be further inspected, e.g., visually and with follow-up observations. In addition to AGEL, which uses a CNN-based search, Huang et al. (2020) apply a residual NN to search through 9000 deg$^2$ from the Dark Energy Camera Legacy Survey (DECaLS) and find 335 strong lensing candidates. However, only with spectroscopic confirmation of the lensing candidates can we verify the sample purity and characteristics to further refine automated searches.

The AGEL survey aims to confirm ~100 bright ($r \lesssim 22$ mag) strong gravitational lenses to enable statistically robust studies of deflectors and magnified sources. Using the J19ab catalogs of lens candidates, we obtain spectroscopic follow-up to measure redshifts for the foreground deflector and background source for lenses that can be observed using telescopes in both hemispheres throughout the year. Most of the arcs and counterimages are at projected distances of $r_{proj} \sim 1''-10''$ and require spatially resolved spectroscopy to separately measure redshifts for both the lenses and the sources.

Building a sample of spectroscopically confirmed strong gravitational lenses opens a range of new discovery space spanning galaxy- to cluster-sized dark matter halos (e.g., Newman et al. 2015; Nord et al. 2016). Confirmed deflectors at $z_{defl} > 0.5$ are especially needed to test for the predicted evolution in mass density profiles with redshift (Sonnenfeld et al. 2013). AGEL also enables the first broad characterization of galaxy populations at source redshifts of $z_{src} \sim 1-4$ at the resolution and signal-to-noise ratio afforded by lensing.

In this paper, we present our first results from the spectroscopic follow-up of the candidate gravitational lenses identified by J19ab in the DES fields and a subsequent search of DECaLS fields using the same method. We summarize how J19ab develop and train the CNN and describe our spectroscopic follow-up with the Keck Observatory and Very Large Telescope (VLT) in Section 2. We discuss our completeness and success rate in confirming strong gravitational lenses in Section 3. We describe the AGEL survey in the context of previous lensing searches and ongoing science analysis in Section 4, and provide our conclusions in Section 5. Unless otherwise noted, we use the AB magnitude system.

## 2. Data and Methods

### 2.1. Convolutional Neural Networks

With advances in computational power and algorithms, we can now expand the boundaries of earlier searches for strong gravitational lenses by applying convolutional neural networks to deep imaging taken by DECam from the Dark Energy Survey (Abbott et al. 2018) and DECaLS (Dey et al. 2019). The coadded DES imaging reaches $r = 24.1$ and has higher angular resolution than SDSS due to a combination of pixel scale ($0\rlap{.}''396$ pix$^{-1}$ versus $0\rlap{.}''263$ pix$^{-1}$) and seeing. CNNs can deliver samples with the highest purity of non-spectroscopic lens-finding algorithms and circumvent a limitation of earlier lens surveys such as SLACS and BELLS that were based on spectroscopic selection with the SDSS fiber (radius of $1\rlap{.}''5$; Bolton et al. 2008; Brownstein et al. 2012). Note that the survey by the Dark Energy Spectroscopic Instrument (DESI) is even more severely limited than SDSS, i.e., the DESI fibers have core diameters of $1\rlap{.}''5$ (Flaugher & Bebek 2014; DESI Collaboration 2016) compared to $3''$ diameter SDSS fibers.

Our sample of lens candidates captures a broader sample of galaxy-scale lenses that includes systems with Einstein radii $> 1\rlap{.}''5$ (see Figure 1). Here we summarize the approach used in J19ab to select gravitational lens candidates in the DES Year 3 and DECaLS DR7 fields and refer the reader to J19ab for a complete description of the CNN method and resulting catalog of candidates.

#### 2.1.1. Training the CNN

Training a CNN to separate lenses and non-lenses requires labeled examples. J19ab used the LensPop code described in Collett (2015) to generate training sets of up to 250,000 images split equally between positive and negative examples; images are each $100 \times 100$ pixels. J19ab trained a CNN with four convolutional layers with different kernel sizes. With each iteration, the loss and accuracy are measured and used to update the weights of the network. Training continued until the validation loss did not improve by more than $10^{-4}$ over six epochs, where a single epoch constitutes one run over the entire training set.

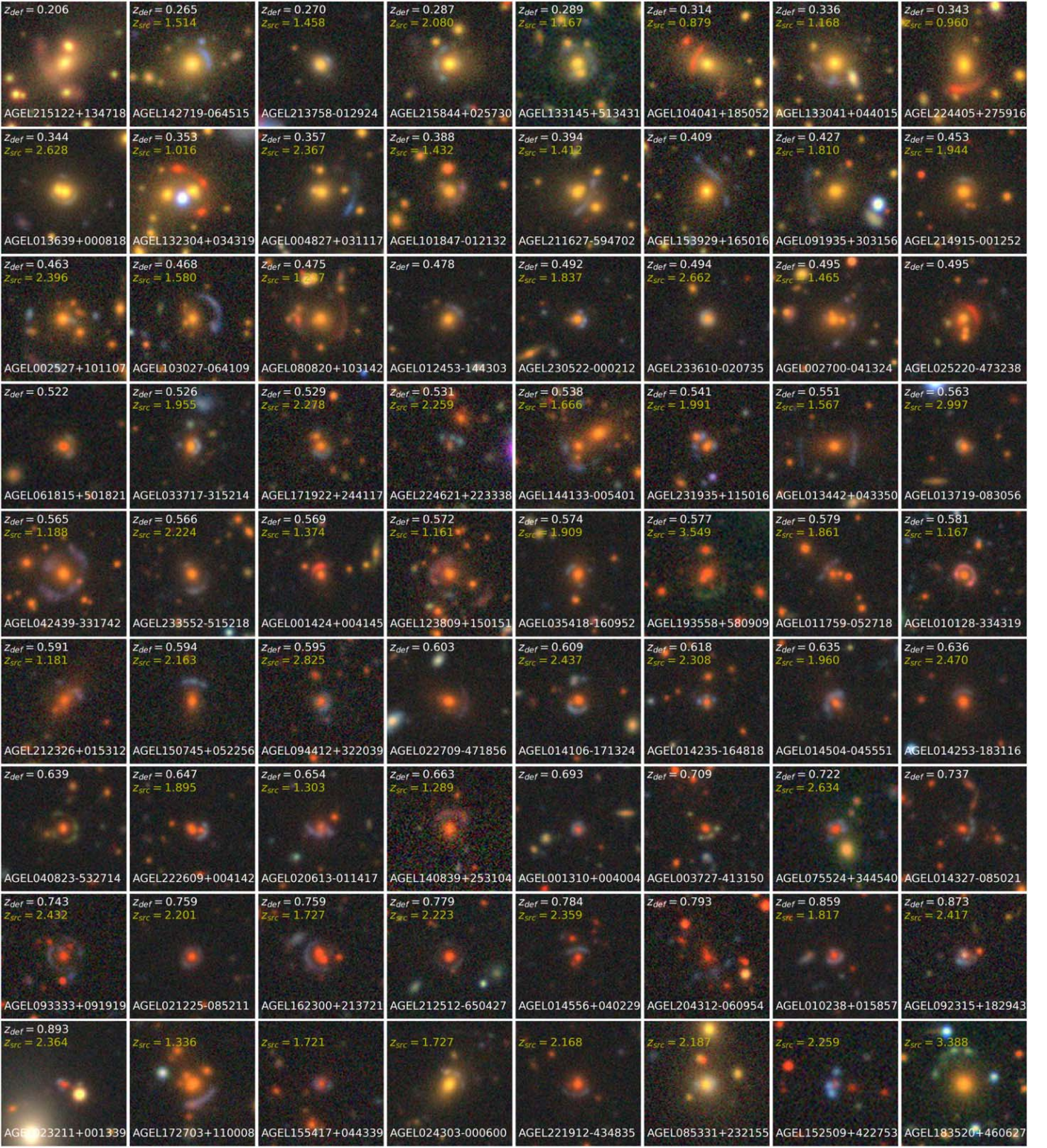

**Figure 1.** Imaging taken by DECam ($26'' \times 26''$) from DES and DECaLS for the gravitational lenses with spectroscopy from our ongoing follow-up campaign combined with published redshifts from the literature (see Section 2.2). Each system has a spectroscopic redshift ($Q_z = 1, 2,$ or $3$) for either the foreground deflector, background source, or both; the lenses are ordered by increasing deflector redshift with the lens candidates that have only source redshifts at the bottom. Considering only redshifts with $Q_z \geqslant 2$, we present 53 confirmed lenses with both $z_{\rm defl}$ and $z_{\rm src}$, and 15 lenses with either $z_{\rm defl}$ or $z_{\rm src}$ (see Table 3 for the AGEL redshift tally). Many systems have Einstein radii larger than the SDSS fiber ($3''$ diameter) and several are likely in clusters/groups.

*2.1.2. Selecting Lens Candidates Using the CNN: DES Year 3 and DECaLS DR7*

The trained CNN was applied to a catalog of approximately 8 million sources from the DES with *gri* photometry to select gravitational lensing candidates. J19ab applied color and magnitude cuts of

$$\{0 < [(g-i) < 3 \bigcap (-0.2)] < [(g-r) < 1.75] \bigcap r \leqslant 22\} \quad (1)$$

to ensure that the sample is not biased against the combined color of the foreground deflector and background source, where the latter tends to be blue at optical wavelengths.

J19ab identified a sample of ∼1300 lens candidates by combining legacy imaging from DES Year 3 and CNNs trained on artificially generated images of lenses (*gri*). All the candidates had been visually inspected and ranked on a 0–3 scale where 0 is not a lens, 1 is possible, 2 is probable, and 3 is definite. J19ab visually examined candidates with lower and lower scores until the purity was only ∼1%, i.e., candidates with lower scores had a likely contamination rate of >99%. The CNNs used by J19ab delivered samples with a purity as high as 20% for the highest scoring candidates, i.e., one in five examined images was a probable or definite lens. We refer the reader to J19ab for a more detailed description of the visual validation process.

To increase sky coverage and take advantage of the DECam Legacy Survey Data Release 7 (DECaLS DR7; Dey et al. 2019), we use the same method from J19ab to identify another ∼600 lens candidates in the DECaLS fields that were observed with the DECam (DR7). The DES fields (5000 deg²) include the South Galactic Cap, while the DECaLS primarily target the SDSS equatorial sky ($-15° < \delta < 34°$). The AGEL catalog of candidate lenses is based on DR7, which includes imaging taken with DECam, MOSAIC-3 on the Mayall telescope, and the 90Prime camera on the Bok telescope. For uniformity of the imaging, note that only the observations from DECaLS taken with DECam are used to select candidate lenses for AGEL.

The CNNs from J19ab were retrained on *grz* imaging from DECaLS DR7 and run on 3.1 million sources. The subset of ∼20,000 most highly scored candidates were then visually inspected by three team experts (C.J., K.G., T.C.). The DECaLS lens candidates were generated separately from the DES candidates and are not published in J19ab. Given the common data sets used to search for gravitational lenses, we note that some of our candidates are in earlier catalogs as well, e.g., Huang et al. (2020, 2021) and Stark et al. (2013).

Figure 3 shows the combined distribution of ∼1900 high-quality candidate gravitational lenses identified in the DES and DECaLS DR7 fields. The candidate gravitational lenses span a range in photometric redshift ($z_{\mathrm{phot}} = 0.39$–$0.81$; see Table 2 and Figure 8) and, due to the magnitude and color cuts, are brighter than $r = 22$ mag (see J19ab; Figure 4).

*2.2. Literature Spectroscopic Redshifts*

In selecting targets for spectroscopic follow-up, we prioritized AGEL candidates with published spectroscopic redshifts for the candidate foreground deflector. However, we did not exclude any lens candidates for spectroscopic follow-up because virtually none had spectroscopic confirmation of both the deflector and source, and independent confirmation is helpful.

**Table 1**
AGEL Spectroscopic Observing Runs (2018 April–2021 March)

| Proposal ID | Telescope/Instrument | Awarded Time | Conditions |
|---|---|---|---|
| 0101.A-0577 | VLT/X-Shooter | 2 nights | clear |
| Engineering time | Keck/NIRES | 1 hr | clear |
| 2019B_W226 | Keck/ESI | 1 night | clear |
| 2019B_U058 | Keck/ESI | 2 nights | clear |
| 2020A_U160 | Keck/NIRES | 1 night | cloudy |
| 2020A_W128 | Keck/ESI | 0.5 night | closed |
| 2020A_U160 | Keck/ESI | 1 night | closed |
| 2020B_U044 | Keck/ESI | 1 night | clear |
| 2020B_U044 | Keck/NIRES | 1 night | mixed |
| 2020B_W127 | Keck/NIRES | 1 night | clear |
| 2021A_U022 | Keck/ESI | 1 night | mixed |
| 2021A_U022 | Keck/NIRES | 1 night | cloudy |
| 2021A_W235 | Keck/NIRES | 1 night | clear |

Of the 79 AGEL spectroscopic targets, 37 have spectroscopic redshifts published in the literature (Table 2). The literature redshifts are from existing surveys that used SDSS observations to select lens candidates including SLACS (131; Bolton et al. 2008), BELLS (45; Brownstein et al. 2012), and CASSOWARY (29; Stark et al. 2013). These surveys combined published redshifts from SDSS and BOSS (Eisenstein et al. 2011) with additional follow-up spectroscopy; we refer the reader to their papers for further detail.

*2.3. Photometry from DESI Legacy Imaging Survey DR9*

In the following analysis, we use updated magnitudes provided by the DESI Legacy Imaging Survey (https://www.legacysurvey.org; Dey et al. 2019) that consolidates photometry taken by multiple telescopes to access both hemispheres. The DESI DR9 provides updated photometry for earlier lens searches including SLACS, BELLS, and CASSOWARY that enables direct comparison to AGEL. DR9 provides total *r* magnitudes (AB system) measured using TRACTOR (for details, see Dey et al. 2019). Every source is modeled using one of six morphological types that is convolved with the specific point-spread function for each exposure (description available on the DR9 website).

We use the total *r*-band magnitude for the DR9 object that is closest in projected distance to the position of the gravitational lens. The gravitational lens is usually centered on the brighter foreground deflector, and the deflector and fainter images of the lensed source are sufficiently separated for our lens candidates such that the reported flux corresponds to the deflector (see Figures 1 and 4). Note that using the ground-based imaging to train our CNN and visual inspection means we are best able to identify lens candidates with $r_{\mathrm{EIN}} \gtrsim 1''$ (Figure 1).

*2.4. Spectroscopy*

The primary goal of our spectroscopic campaign is to confirm as many gravitational lens candidates from the CNN-selected catalogs as possible. To secure spectroscopic confirmation of the candidate gravitational lenses, we use observations from the Keck Observatory and Very Large Telescope for northern and southern targets respectively. Observations were carried over 13.5 nights from 2018 April to 2021 March with varying conditions including telescope closures due to the 2020 pandemic (Table 1). Whether candidate sources or deflectors were targeted depended on the instrument: the key spectral features of deflectors ($z_{\rm defl} < 1$) are captured with optical wavelength coverage by Keck/ESI while the higher-redshift sources ($z_{\rm src} > 1$) are better matched to near-IR (NIR) wavelength coverage by Keck/NIRES. Both deflectors and sources can be confirmed with VLT/X-Shooter with continuous optical–NIR coverage.

Targets were selected to be visible during awarded nights at optimal airmasses and have Einstein radius $r_{\rm EIN} \gtrsim 1''$ (Figures 1 and 2). Higher priority was given to targets that (1) were near suitable guide-stars for future follow-up with adaptive optics, (2) have imaging with the Hubble Space Telescope, and/or (3) had previously known spectroscopic redshifts of the candidate deflector that enable efficient confirmation of background arcs. No other criteria were used to prioritize the lens targets. Our general strategy was to target the single brightest arc and the candidate deflector.

We focus on the spectroscopic redshifts for the analysis in this paper. We note that the spectra are of sufficient quality to measure velocity dispersions for the foreground deflectors and gas kinematics in the lensed sources (G. C. Vasan et al 2022, in preparation).

### 2.4.1. Keck Spectroscopy

We use the Echellette Spectrograph and Imager (ESI, Sheinis et al. 2002) and NIRES (Wilson et al. 2004) instruments on the Keck telescopes to obtain optical and near-infrared spectroscopy respectively of the candidate gravitational lenses (Table 1). ESI was used primarily to measure redshifts for the foreground deflectors ($z_{\rm defl} \lesssim 1$). Depending on the redshift of the source, a spectroscopic redshift could be obtained with ESI via interstellar medium (ISM) absorption lines or Ly$\alpha$ emission, or with NIRES via emission lines.

With ESI in echelle mode (slit length of $20''$), we obtain spectroscopy at 3900–10900 Å with a corresponding dispersion of 0.16–0.30 Å pix$^{-1}$ from order 15 to 30. We use a slit width of $1''\!.0$, providing a resolving power of $R = 4000$, and typical total exposure time on target of 20–80 minutes depending on conditions. The ESI data are reduced using the ESIRedux (2019 runs) and makee (2020 and 2021 runs; see Table 1) pipelines provided by J. X. Prochaska[10] and T. Barlow[11] respectively.

We use NIRES primarily to target the background sources at higher redshifts ($z_{\rm src} > 1$) because sources tend to be star-forming galaxies with emission lines. Using the fixed slit width of $0''\!.55$, the wavelength coverage is 0.9–2.45 $\mu$m with a mean spectral resolution of 2700 and spectrometer pixel scale of $0''\!.15$ pix$^{-1}$. The NIRES slit length is $18''$ and typical dither steps are $\pm (3-7)''$. The typical total exposure time on target was 20 minutes (ABBA dither pattern) and the data were reduced using the NSX pipeline written by T. Barlow.[12]

The redshift precisions of ESI and NIRES are comparable given the pixel scales and spectral resolutions. Spectra from both instruments can be flux-calibrated using a standard star taken during the respective observing runs. However, flux calibration is not needed for the redshift confirmations that are the focus on this paper. For the same reason, the spectra have not been corrected for telluric absorption.

---

[10] ESIRedux pipeline
[11] makee pipeline
[12] NSX pipeline

### 2.4.2. Very Large Telescope Spectroscopy

We use the ESO/VLT X-Shooter instrument (Vernet et al. 2011) to obtain spectroscopy at 3000–25000 Å (Table 1). We use slit widths of $1''\!.0$, $0''\!.9$, and $0''\!.9$ with corresponding spectral resolutions of 5400, 8900, and 5600 for the UVB (300–560 nm), VIS (560–1024 nm), and NIR (1024–2480 nm) arms respectively. The typical total exposure time is 10–40 minutes on the deflectors and 40–60 minutes on the lensed sources. In some cases a single slit is placed across the deflector and lensed sources, and in other cases separate slit positions are used because of the lens geometry. The data are reduced using the REFLEX pipeline provided by ESO (Modigliani et al. 2010) and publicly available 2D to 1D extraction code from Corentin Schreiber.[13]

The X-Shooter spectra are flux-calibrated using a standard star taken during the respective observing runs. However, flux calibration is not needed for the redshift confirmations that are the focus on this paper. For the same reason, the spectra have not been corrected for telluric absorption.

### 2.5. Determining Spectroscopic Redshifts

Spectroscopy is essential for determining accurate redshifts of the targeted systems, especially for gravitational lenses where blended light from multiple objects makes obtaining photometric redshifts for sources challenging. The spectra are reduced using their respective instrument pipelines that perform bias, dark current, cosmic ray, and sky subtraction, and flat-field corrections. The 2D spectra are fit along the slit (spatial) axis with a Gaussian profile, and the 1D spectra are extracted from the 3$\sigma$ region of the fitted Gaussian.

Precise spectroscopic redshifts are determined by using a custom Python script to fit Gaussians to the emission and absorption lines in the 1D spectra. A subset of the targets (30/79) have a photometric redshift for the foreground deflector from existing public catalogs, and we use $z_{\rm phot}$ for the initial guess to determine the spectroscopic redshift. Note that like all ground-based spectroscopic surveys, we are incomplete at certain redshifts due to spectral features falling in optical/NIR bands of atmospheric absorption.

None of the background sources have photometric redshifts because the images of the lensed sources are faint and frequently blended, e.g., with the foreground galaxies. As we discuss in Section 3.3, the photometric redshifts for the foreground deflectors are remarkably reliable despite potentially blended photometry. However, follow-up spectroscopy of both candidate deflector and source is essential to confirm whether the gravitational lens is real (see Section 3.1).

Depending on the redshift of the object and wavelength coverage (optical versus NIR), we use different spectral features in the 1D spectra to measure redshifts (Figure 5). Source redshifts measured with NIRES and X-Shooter are almost exclusively determined using rest-frame optical emission lines. Source redshifts from ESI are mostly from interstellar absorption lines except for sources at $z_{\rm src} < 1.7$, where the redshifts are mostly from [O II]. Note that redshifts from ISM absorption lines are not systemic and are typically blueshifted by $\sim 200$ km s$^{-1}$ due to, e.g., large-scale outflows (Rakic et al. 2011).

---

[13] Available on github

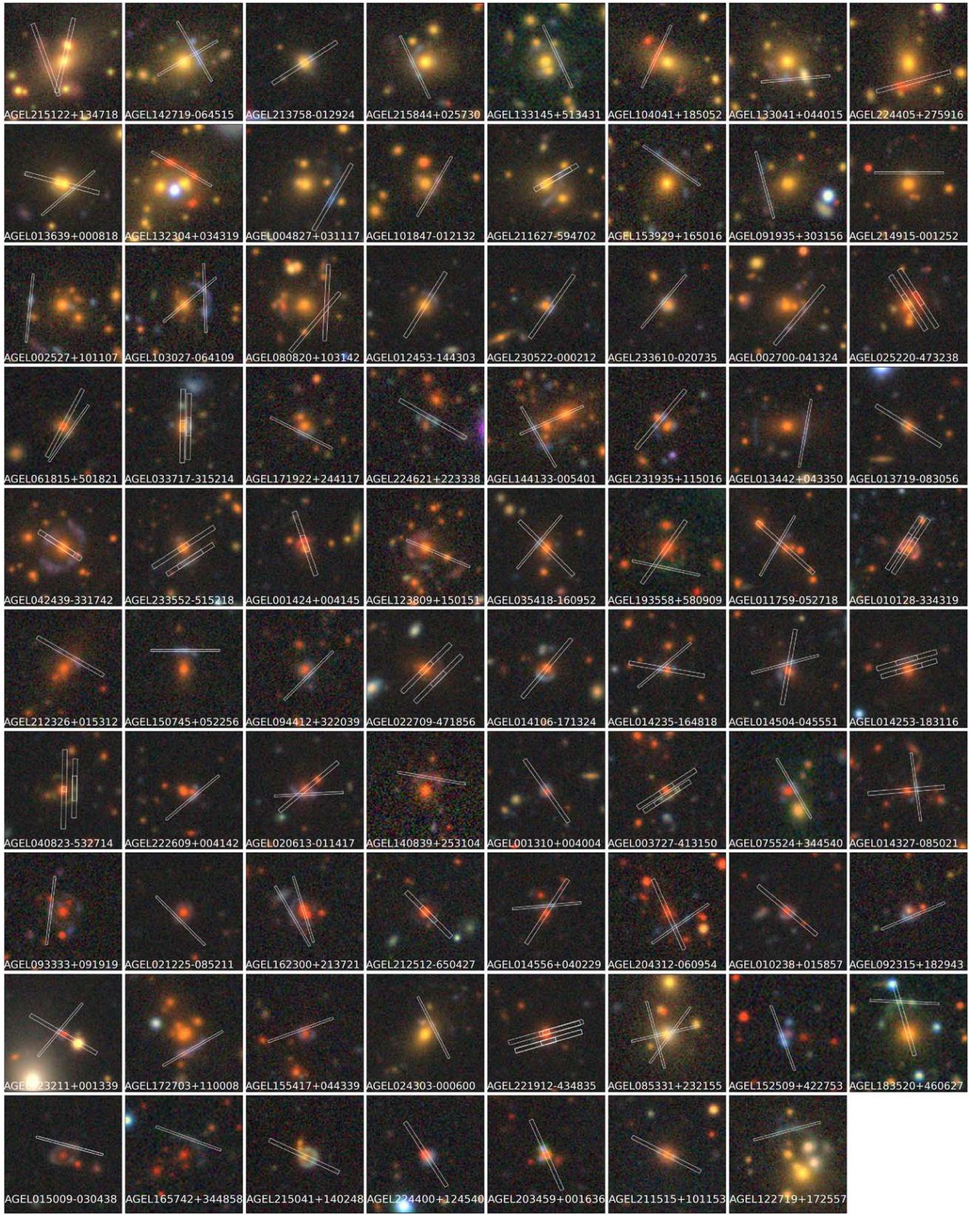

**Figure 2.** The slit positions of our spectroscopic observations overlaid on the DECam imaging ($26'' \times 26''$) where the observed lens candidates are in the same order as in Figure 1 (north up, east to the left). The bottom row of seven systems are confirmed either to not be a lens or to have $Q_z \leqslant 1$ (Table 2). We select spectroscopic targets from the CNN-selected catalogs described in Section 2.1, where systems with spectroscopic redshifts from the literature are prioritized (see Section 2.2). Considering only spectroscopic redshifts with $Q_z \geqslant 2$, the AGEL redshift tally (Table 3) includes 53 confirmed lenses with both $z_{\text{defl}}$ and $z_{\text{src}}$, and 15 lenses with either $z_{\text{defl}}$ or $z_{\text{src}}$ (Table 2).

For the foreground deflectors ($z_{\rm defl} < 1$), spectral features measured by ESI are mostly absorption lines including H$\beta$, H$\gamma$, the H & K calcium lines, and Mgb. The spectral features are easily identified with visual inspection to obtain an initial estimate for determining a spectroscopic redshift. For the background sources ($z_{\rm src} > 1$), spectral features measured with ESI, NIRES, or X-Shooter are usually emission lines including [O II] $\lambda 3727$, H$\beta$, H$\alpha$, and [O III] $\lambda 5007$. For higher-redshift sources ($z_{\rm src} \gtrsim 2$), we sometimes obtain UV absorption lines including C IV $\lambda 1550$, Fe II $\lambda 1608$, and Al II $\lambda 1670$ with ESI.

Using multiple spectral features (see Figure 5) results in low redshift uncertainties of <0.00005. The spectral lines have minimum signal-to-noise ratio (S/N) of 3 and we visually inspected all of the Gaussian fits. The spectral templates provided the initial redshift for fitting, and the spectroscopic redshift is the median redshift measured using multiple lines. For the sources, the spectral lines are boosted by the gravitational lensing.

The 106 objects targeted for spectroscopic follow-up are listed in Table 2. As described earlier in Section 2.4, candidate deflectors and sources are selected based on the instrument (optical and/or NIR coverage) with the goal of obtaining spectroscopic redshifts for the deflector and at least one lensed image of the source. To quantify the robustness of each spectroscopic redshift, we assign a redshift quality flag $Q_z$ by inspecting and comparing their 1D and, where available, 2D spectra. Following Tran et al. (2015), a quality flag of:

1. $Q_z = 3$ denotes a robust measurement (multiple spectral lines). Includes resolved [O II] doublet with S/N $\geqslant 3$.
2. $Q_z = 2$ is likely (single spectral line with potential secondary line).
3. $Q_z = 1$ is a guess (single line and/or no strong spectral features).

The spectra shown in Figure 5 all have $Q_z = 3$. For comparison, spectra with $Q_z < 3$ are shown in Figure 6. We measure 104 redshifts, but in our analysis we consider only the 95 redshifts with $Q_z \geqslant 2$ (Figure 7).

## 3. Results

### 3.1. Spectroscopic Success Rate with CNN-based Search

During the observing runs listed in Table 1, we targeted 106 objects in 79 candidate gravitational lenses for spectroscopy and measure 104 redshifts. We are unable to measure a redshift for two of the targets due to lack of spectral features (Table 3); in some cases, the spectral features may fall in windows of atmospheric absorption. We define the spectroscopic success rate as the ratio of 104 redshifts to 106 targets, which is 98% (Table 3).

Of the 79 candidate gravitational lenses that we targeted, we obtain redshifts in 77 systems. Our spectroscopy confirms that one object is a (red) Milky Way M-star and three are galaxies at $z_{\rm spec} < 0.5$ (Table 3). The three galaxies are a rotating ring galaxy (AGEL 215041+140248), a rotating ring galaxy where the "arc" is part of the ring (AGEL 211515+101153), and a system where the "arc" and "deflector" are at the same redshift (AGEL 224400+124540). Removing these four systems from our analysis leaves 73 gravitational lenses where we secure redshifts for either the foreground deflector, the background source, or both (Table 3). We then apply a redshift quality requirement of $Q_z \geqslant 2$ that removes five candidate lenses.

In the following analysis, we use only the 68 strong lensing systems that satisfy these criteria: (1) not spectroscopically confirmed to be a star or multicomponent galaxy at $z_{\rm spec} < 0.5$; (2) spectroscopic redshifts for the foreground deflector and/or background source; (3) spectroscopic redshifts with $Q_z \geqslant 2$; and (4) if both $z_{\rm defl}$ and $z_{\rm src}$ are measured, $z_{\rm src} > z_{\rm defl}$. Of the 68 strong lenses, 53 have $z_{\rm defl}$ and $z_{\rm src}$ from a combination of our spectroscopic follow-up and published values in the literature, and 15 have either $z_{\rm defl}$ or $z_{\rm src}$ (Tables 2 and 3; Figures 1 and 2).

For the seven systems where we have spectroscopic redshifts for deflectors from our own observations as well as values from the literature, we use our redshifts. The spectroscopic redshifts are consistent: the median absolute difference for these seven deflectors is 0.001 with semi-interquartile range of 0.028. The two largest outliers are at $z_{\rm spec} \sim 0.7$ (see Table 2).

Our results establish that CNN-based search methods are highly effective at identifying strong gravitational lenses in imaging and strongly support using CNNs in future surveys by LSST and EUCLID. We confirm a high success rate of 88% for the CNN-selected candidates by taking the ratio of the 68 strong lenses to the total number of 77 systems with measured redshifts (Table 3). The 88% is likely a conservative lower limit: if we exclude only the four non-lenses and relax the spectroscopic quality flag to use the remaining 73, the confirmation rate of CNN-selected lens candidates is 95%.

### 3.2. Spectroscopic Redshifts of Gravitational Lenses

Of the 68 gravitational lenses with secure redshifts ($Q_z \geqslant 2$), 53 have spectroscopic redshifts for both the foreground deflector and background source (Table 2). The spectroscopic redshifts for 25 of the deflectors are from our spectroscopy, and 28 are published redshifts from surveys including BOSS (Eisenstein et al. 2011) and CASSOWARY (Stark et al. 2013). For the 53 confirmed gravitational lenses, the average redshifts and standard deviations for the deflectors and sources are $0.55 \pm 0.15$ and $1.91 \pm 0.47$ respectively (Table 4).

We include 15 systems with a spectroscopic redshift for either the candidate foreground deflector or background source (but not both; see Table 2), and we are continuing our spectroscopic follow-up of these 15 systems. We are confident that the 15 strong lenses are real given our statistics, existing high-resolution Hubble Space Telescope imaging for a subset, and their spectroscopic redshift distributions. We have obtained redshifts for eight deflectors ($0.21 \leqslant z_{\rm defl} \leqslant 0.79$) in the 15 systems, and seven are at $z_{\rm defl} \geqslant 0.5$. For seven of the 15 systems, we have redshifts for the sources confirming they are at $z_{\rm spec} = 1.336$–$3.388$. For comparison, the four systems that we confirm to not be lenses are all at $z_{\rm spec} < 0.5$.

Our results confirm that existing imaging surveys are able to detect strongly lensed sources at $z_{\rm src} \gtrsim 2$. Included are 41 deflectors at $z_{\rm defl} > 0.5$ that are especially useful for measuring how mass density profiles evolve with redshift (see Figure 7). Considering systems where we have secured a redshift for the deflector or source (Table 2), we have 35 deflectors and 60 sources with average redshifts of $0.58 \pm 0.14$ and $1.92 \pm 0.59$ respectively (Figure 7).

### 3.3. Precision of Photometric Redshifts

We have spectroscopic redshifts for 23 systems with photometric redshifts for the foreground deflectors determined using DES and DECaLS photometry. For the DES lens

Table 2
AGEL Survey Spectroscopic Redshifts

| AGEL ID[a] | $\alpha_{J2000}$ | $\delta_{J2000}$ | $r$ mag[b] | $z_{phot}$ | $z_{Lit}$[c] | Literature[d] | $z_{defl}$[e] | $Q_{zdefl}$[f] | $z_{src}$[e] | $Q_{zsrc}$[f] |
|---|---|---|---|---|---|---|---|---|---|---|
| AGEL 001310+004004 | 00:13:09.6 | 00:40:03.6 | 20.573 | 0.750 | 0.693 | BOSS | 0.69325 | 3 | 2.07000 | 1 |
| AGEL 001424+004145 | 00:14:24.3 | 00:41:45.5 | 20.039 | 0.590 | 0.570 | BOSS | 0.56850 | 3 | 1.37389 | 2 |
| AGEL 002527+101107 | 00:25:27.4 | 10:11:07.1 | 18.294 | ⋯ | 0.463 | BOSS | ⋯ | ⋯ | 2.39628 | 3 |
| AGEL 002700-041324 | 00:27:00.1 | −04:13:23.6 | 18.406 | 0.570 | 0.495 | BOSS | ⋯ | ⋯ | 1.46470 | 2 |
| AGEL 003727-413150 | 00:37:27.1 | −41:31:49.8 | 20.768 | 0.670 | ⋯ | ⋯ | 0.70884 | 3 | ⋯ | ⋯ |
| AGEL 004827+031117 | 00:48:27.2 | 03:11:17.1 | 18.253 | 0.390 | 0.357 | BOSS | ⋯ | ⋯ | 2.36671 | 3 |
| AGEL 010128-334319 | 01:01:27.8 | −33:43:19.2 | 19.274 | 0.630 | ⋯ | ⋯ | 0.58110 | 3 | 1.16674 | 3 |
| AGEL 010238+015857 | 01:02:38.3 | 01:58:56.7 | 20.663 | 0.810 | 0.869 | BOSS | ⋯ | ⋯ | 1.81696 | 3 |
| AGEL 011759-052718 | 01:17:58.7 | −05:27:17.7 | 19.609 | 0.570 | 0.580 | BOSS | 0.57916 | 3 | 1.86051 | 2 |
| AGEL 012453-144303 | 01:24:53.1 | −14:43:02.6 | 18.918 | 0.460 | ⋯ | ⋯ | 0.47781 | 3 | ⋯ | ⋯ |
| AGEL 013442+043350 | 01:34:42.4 | 04:33:50.0 | 18.778 | ⋯ | 0.551 | BOSS | ⋯ | ⋯ | 1.56704 | 3 |
| AGEL 013639+000818 | 01:36:39.2 | 00:08:18.1 | 17.861 | ⋯ | 0.344 | SDSS legacy | 0.34405 | 3 | 2.62833 | 3 |
| AGEL 013719-083056 | 01:37:18.8 | −08:30:55.9 | 19.896 | 0.510 | ⋯ | ⋯ | 0.56300 | 3 | 2.99700 | 3 |
| AGEL 014106-171324 | 01:41:06.1 | −17:13:23.7 | 19.747 | 0.610 | ⋯ | ⋯ | 0.60873 | 3 | 2.43700 | 2 |
| AGEL 014235-164818 | 01:42:35.0 | −16:48:17.5 | 20.024 | ⋯ | ⋯ | ⋯ | 0.61781 | 3 | 2.30792 | 3 |
| AGEL 014253-183116 | 01:42:52.9 | −18:31:15.8 | 19.690 | 0.690 | ⋯ | ⋯ | 0.63627 | 3 | 2.46972 | 3 |
| AGEL 014327-085021 | 01:43:26.9 | −08:50:21.3 | 20.812 | 0.680 | 0.680 | ⋯ | 0.73701 | 3 | 2.75500 | 1 |
| AGEL 014504-045551 | 01:45:04.3 | −04:55:51.0 | 19.451 | 0.600 | ⋯ | CASSOWARY103 | 0.63536 | 3 | 1.95963 | 3 |
| AGEL 014556+040229 | 01:45:56.3 | 04:02:29.0 | 20.655 | 0.660 | ⋯ | ⋯ | 0.78390 | 3 | 2.35921 | 3 |
| AGEL 015009-030438 | 01:50:09.1 | −03:04:38.3 | 21.226 | 0.680 | ⋯ | ⋯ | 0.63675 | 1 | ⋯ | ⋯ |
| AGEL 020613-011417 | 02:06:13.5 | −01:14:17.4 | 19.988 | 0.770 | 0.714 | BOSS | 0.65402 | 2 | 1.30323 | 3 |
| AGEL 021225-085211 | 02:12:25.2 | −08:52:10.8 | 20.282 | 0.690 | 0.759 | BOSS | ⋯ | ⋯ | 2.20096 | 2 |
| AGEL 022709-471856 | 02:27:09.0 | −47:18:55.8 | 19.078 | 0.680 | ⋯ | ⋯ | 0.60290 | 3 | ⋯ | ⋯ |
| AGEL 023211+001339 | 02:32:11.2 | 00:13:39.2 | 21.620 | 0.810 | ⋯ | ⋯ | 0.89293 | 2 | 2.36441 | 3 |
| AGEL 024303-000600 | 02:43:03.0 | −00:06:00.2 | 18.095 | ⋯ | ⋯ | ⋯ | ⋯ | ⋯ | 1.72698 | 3 |
| AGEL 025220-473238 | 02:52:19.9 | −47:32:37.7 | 18.524 | 0.490 | ⋯ | ⋯ | 0.49515 | 3 | ⋯ | ⋯ |
| AGEL 033717-315214 | 03:37:17.2 | −31:52:13.6 | 19.170 | 0.470 | ⋯ | ⋯ | 0.52560 | 3 | 1.95476 | 3 |
| AGEL 035418-160952 | 03:54:18.3 | −16:09:52.2 | 19.222 | 0.630 | ⋯ | ⋯ | 0.57407 | 3 | 1.90925 | 3 |
| AGEL 040823-532714 | 04:08:22.7 | −53:27:14.2 | 20.317 | ⋯ | ⋯ | ⋯ | 0.63933 | 2 | ⋯ | ⋯ |
| AGEL 042439-331742 | 04:24:38.7 | −33:17:41.7 | 18.194 | 0.620 | ⋯ | ⋯ | 0.56491 | 3 | 1.18842 | 3 |
| AGEL 061815+501821 | 06:18:15.3 | 50:18:21.2 | 19.177 | ⋯ | ⋯ | ⋯ | 0.52207 | 3 | ⋯ | ⋯ |
| AGEL 075524+344540 | 07:55:23.5 | 34:45:39.6 | 20.533 | ⋯ | 0.722 | BOSS | ⋯ | ⋯ | 2.63350 | 2 |
| AGEL 080820+103142 | 08:08:20.4 | 10:31:42.2 | 18.056 | ⋯ | 0.475 | BOSS | ⋯ | ⋯ | 1.23742 | 2 |
| AGEL 085331+232155 | 08:53:31.2 | 23:21:54.7 | 17.809 | ⋯ | ⋯ | ⋯ | ⋯ | ⋯ | 2.18739 | 3 |
| AGEL 091935+303156 | 09:19:35.0 | 30:31:56.3 | 18.133 | ⋯ | 0.427 | BOSS | ⋯ | ⋯ | 1.81039 | 3 |
| AGEL 092315+182943 | 09:23:14.6 | 18:29:43.4 | 19.443 | ⋯ | 0.873 | BOSS | ⋯ | ⋯ | 2.41673 | 3 |
| AGEL 093333+091919 | 09:33:33.3 | 09:19:19.0 | 20.632 | ⋯ | 0.743 | BOSS | ⋯ | ⋯ | 2.43243 | 2 |
| AGEL 094412+322039 | 09:44:11.8 | 32:20:38.8 | 19.954 | ⋯ | 0.595 | BOSS | ⋯ | ⋯ | 2.82512 | 3 |
| AGEL 101847-012132 | 10:18:47.3 | −01:21:32.6 | 18.384 | ⋯ | 0.388 | BOSS | ⋯ | ⋯ | 1.43210 | 3 |
| AGEL 103027-064109 | 10:30:27.2 | −06:41:08.9 | 18.631 | ⋯ | ⋯ | ⋯ | 0.46775 | 3 | 1.58017 | 3 |
| AGEL 104041+185052 | 10:40:41.2 | 18:50:51.7 | 16.904 | ⋯ | 0.314 | SDSS Legacy | ⋯ | ⋯ | 0.87872 | 3 |
| AGEL 122719+172557 | 12:27:19.0 | +17:25:56.6 | ⋯ | ⋯ | ⋯ | ⋯ | ⋯ | ⋯ | ⋯ | ⋯ |
| AGEL 123809+150151 | 12:38:08.9 | 15:01:51.2 | 18.378 | ⋯ | ⋯ | ⋯ | 0.57160 | 3 | 1.16149 | 3 |
| AGEL 132304+034319 | 13:23:04.1 | 03:43:19.4 | 17.507 | ⋯ | 0.353 | BOSS | ⋯ | ⋯ | 1.01590 | 3 |
| AGEL 133041+044015 | 13:30:40.6 | 04:40:14.5 | 17.149 | ⋯ | 0.336 | SDSS Legacy | ⋯ | ⋯ | 1.16757 | 3 |
| AGEL 133145+513431 | 13:31:45.3 | 51:34:31.1 | 17.504 | ⋯ | 0.289 | BOSS | ⋯ | ⋯ | 1.16749 | 3 |
| AGEL 140839+253104 | 14:08:38.7 | 25:31:04.0 | 19.595 | ⋯ | 0.663 | BOSS | ⋯ | ⋯ | 1.28944 | 3 |
| AGEL 142719-064515 | 14:27:18.7 | −06:45:14.9 | 16.790 | ⋯ | ⋯ | ⋯ | 0.26500 | 3 | 1.51409 | 3 |
| AGEL 144133-005401 | 14:41:33.0 | −00:54:01.4 | 18.082 | ⋯ | ⋯ | ⋯ | 0.53761 | 3 | 1.66569 | 3 |
| AGEL 150745+052256 | 15:07:45.1 | 05:22:56.3 | 19.530 | ⋯ | 0.594 | BOSS | ⋯ | ⋯ | 2.16275 | 3 |
| AGEL 152509+422753 | 15:25:09.0 | 42:27:52.6 | 22.011 | ⋯ | ⋯ | ⋯ | ⋯ | ⋯ | 2.25890 | 3 |
| AGEL 153929+165016 | 15:39:29.0 | 16:50:16.4 | 17.863 | ⋯ | 0.409 | BOSS | ⋯ | ⋯ | ⋯ | ⋯ |
| AGEL 155417+044339 | 15:54:16.6 | 04:43:39.2 | 20.868 | ⋯ | ⋯ | ⋯ | ⋯ | ⋯ | 1.72064 | 2 |
| AGEL 162300+213721 | 16:23:00.3 | 21:37:21.4 | 20.271 | ⋯ | ⋯ | ⋯ | 0.75921 | 3 | 1.72698 | 3 |
| AGEL 165742+344858 | 16:57:41.6 | 34:48:58.3 | 21.165 | ⋯ | ⋯ | ⋯ | ⋯ | ⋯ | 2.46382 | 1 |
| AGEL 171922+244117 | 17:19:21.5 | 24:41:16.7 | 19.409 | ⋯ | 0.529 | BOSS | ⋯ | ⋯ | 2.27766 | 3 |
| AGEL 172703+110008 | 17:27:03.3 | 11:00:07.6 | 18.569 | ⋯ | ⋯ | ⋯ | ⋯ | ⋯ | 1.33634 | 2 |
| AGEL 183520+460627 | 18:35:20.1 | 46:06:27.4 | 18.314 | ⋯ | ⋯ | ⋯ | ⋯ | ⋯ | 3.38845 | 3 |
| AGEL 193558+580909 | 19:35:58.2 | 58:09:09.0 | 19.355 | ⋯ | ⋯ | ⋯ | 0.57744 | 3 | 3.54898 | 3 |
| AGEL 203459+001636 | 20:34:58.6 | 00:16:35.5 | 19.839 | ⋯ | ⋯ | MW Mstar | [0] | | | |
| AGEL 204312-060954 | 20:43:12.5 | −06:09:53.6 | 20.343 | ⋯ | ⋯ | ⋯ | 0.79261 | 3 | ⋯ | ⋯ |
| AGEL 211515+101153 | 21:15:15.1 | 10:11:53.1 | 19.227 | ⋯ | ⋯ | low-z galaxy | [0.25] | | | |
| AGEL 211627-594702 | 21:16:27.3 | −59:47:01.8 | 17.892 | 0.480 | ⋯ | ⋯ | 0.39365 | 3 | 1.41166 | 3 |
| AGEL 212326+015312 | 21:23:26.0 | 01:53:12.1 | 19.194 | ⋯ | 0.591 | BOSS | ⋯ | ⋯ | 1.18096 | 3 |

**Table 2**
(Continued)

| AGEL ID[a] | $\alpha_{J2000}$ | $\delta_{J2000}$ | $r$ mag[b] | $z_{phot}$ | $z_{Lit}$[c] | Literature[d] | $z_{defl}$[e] | $Q_{zdefl}$[f] | $z_{src}$[e] | $Q_{zsrc}$[f] |
|---|---|---|---|---|---|---|---|---|---|---|
| AGEL 212512−650427 | 21:25:12.0 | −65:04:26.7 | 20.423 | 0.780 | ⋯ | ⋯ | 0.77900 | 3 | 2.22253 | 2 |
| AGEL 213758−012924 | 21:37:58.0 | −01:29:23.9 | 18.410 | 0.410 | 0.270 | BOSS | ⋯ | ⋯ | 1.45820 | 1 |
| AGEL 214915−001252 | 21:49:15.3 | −00:12:51.5 | 18.164 | 0.640 | 0.453 | BOSS | ⋯ | ⋯ | 1.94381 | 3 |
| AGEL 215041+140248 | 21:50:41.1 | 14:02:48.1 | 19.273 | ⋯ | ⋯ | ring galaxy | [0.481] | ⋯ | | |
| AGEL 215122+134718 | 21:51:21.8 | 13:47:18.2 | 16.638 | ⋯ | ⋯ | ⋯ | 0.20643 | 3 | ⋯ | ⋯ |
| AGEL 215844+025730 | 21:58:43.7 | 02:57:30.2 | 17.287 | ⋯ | 0.287 | BOSS | ⋯ | ⋯ | 2.08015 | 3 |
| AGEL 221912−434835 | 22:19:12.4 | −43:48:35.1 | 19.875 | 0.710 | ⋯ | ⋯ | ⋯ | ⋯ | 2.16767 | 3 |
| AGEL 222609+004142 | 22:26:09.3 | 00:41:42.1 | 20.601 | ⋯ | 0.647 | BOSS | ⋯ | ⋯ | 1.89497 | 3 |
| AGEL 224400+124540 | 22:44:00.3 | 12:45:39.6 | 19.881 | ⋯ | ⋯ | nearby galaxy | [0.078] | ⋯ | | |
| AGEL 224405+275916 | 22:44:04.9 | 27:59:15.7 | 17.554 | ⋯ | 0.343 | BOSS | ⋯ | ⋯ | 0.96034 | 3 |
| AGEL 224621+223338 | 22:46:21.2 | 22:33:37.6 | 21.492 | ⋯ | 0.531 | BOSS | ⋯ | ⋯ | 2.25900 | 3 |
| AGEL 230522−000212 | 23:05:21.7 | −00:02:11.7 | 19.408 | ⋯ | 0.492 | BOSS | ⋯ | ⋯ | 1.83700 | 2 |
| AGEL 231935+115016 | 23:19:34.5 | 11:50:15.9 | 20.233 | ⋯ | 0.540 | BOSS | 0.54122 | 3 | 1.99099 | 3 |
| AGEL 233552−515218 | 23:35:51.9 | −51:52:17.8 | 19.071 | 0.610 | ⋯ | ⋯ | 0.56600 | 3 | 2.22450 | 3 |
| AGEL 233610−020735 | 23:36:10.3 | −02:07:35.0 | ⋯ | ⋯ | 0.494 | BOSS | ⋯ | ⋯ | 2.66173 | 3 |

**Notes.**
[a] AGEL designated identification.
[b] AB magnitudes are from the DESI Legacy Imaging Surveys Data Release 9 (Dey et al. 2019).
[c] Spectroscopic redshift available in the literature including from BOSS, SLACS, BELLS, and CASSOWARY (Bolton et al. 2008; Eisenstein et al. 2011; Brownstein et al. 2012; Stark et al. 2013).
[d] Literature source for $z_{Lit}$.
[e] Typical uncertainty in the spectroscopic redshifts for the deflectors and the sources is $\delta(z) < 0.00005$. Spectroscopic redshifts with [ ] denote systems that are not lenses.
[f] Redshift quality flag where $Q_z$ values of (3, 2, 1) correspond to (robust, probable, guess). In our analysis, we focus on redshifts with $Q_z \geqslant 2$.

**Table 3**
AGEL Spectroscopic Statistics

| Category | Ratio | Fraction |
|---|---|---|
| Spectroscopic targets | 106/106 | 100% |
| Spectroscopic redshifts | 104/106 | 98% |
| Spectroscopic redshifts $Q_z \geqslant 2$ | 95/106 | 90% |
| Gravitational lens candidates | 79/79 | 100% |
| Gravitational lens redshifts | 77/79 | 98% |
| $z_{defl}$[a] OR $z_{src}$ | 73/77 | 95% |
| $z_{defl}$[a] OR $z_{src}$ AND $Q_z \geqslant 2$ | 68/77 | 88% |
| CNN success rate | 68/77 | 88% |
| $z_{defl}$[a] | 61/77 | 80% |
| $z_{src}$ | 60/77 | 78% |
| $z_{defl}$[a] AND $z_{src}$ AND $Q_z \geqslant 2$ | 53/77 | 75% |
| Not lenses | 4/77 | 5% |

**Note.**
[a] For the deflector, we supplement our spectroscopic redshifts with measurements in the literature (see Section 2.2).

candidates, J19ab estimated $z_{phot}$ for the deflectors using the BPZ code (Benítez 2000) with the 3″ aperture $gri$ photometry (Abbott et al. 2018). For DECaLS systems, we used the BPZ code on the model $grz$ photometry published in the DECaLS catalogs (Dey et al. 2019). Due to the spatial resolution of the ground-based DES and DECaLS imaging, our lens candidates tend to be "wide-angle" systems ($r_{EIN} \gtrsim 1″$; Figure 1).

We find the photometric redshifts are remarkably consistent with the spectroscopic values (see Figure 8). The average photometric and spectroscopic redshifts are $\langle z_{phot} \rangle = 0.63 \pm 0.10$ and $\langle z_{spec} \rangle = 0.62 \pm 0.11$. The average absolute difference is $\langle |\delta z|/(1 + z_{spec}) \rangle = 0.03 \pm 0.02$. There is no systematic offset in deflector redshift for lenses with spectroscopic redshifts from BOSS compared to those without.

Our results indicate that potentially blended light from the background source has minimal impact on determining a photometric redshift for the foreground deflector for our sample of lenses ($r_{EIN} \gtrsim 1″$). Because the images of the higher-redshift sources are fainter and can be blended with the foreground deflector, photometric redshifts are not available for the sources. Thus only with follow-up spectroscopy can we confirm whether a system is a true gravitational lens by obtaining spectroscopic redshifts for both the foreground deflector and higher-redshift source.

### 4. Discussion

#### 4.1. Comparison to Previous Lensing Searches

AGEL has now confirmed more strong gravitational lenses than any single previous survey except for SLACS (Table 4). AGEL's key advantages for pushing to higher redshifts than previous searches are the deeper and higher-resolution imaging from DECam, and spectroscopy spanning optical to near-infrared wavelengths. Notably, the 68 AGEL systems have a higher average deflector redshift ($0.58 \pm 0.14$) than many previous surveys including SLACS, BELLS, CASSOWARY, and SL2S (Figures 7 and 9; Table 4). The average spectroscopic redshifts for the foreground deflectors in the fiber searches by SLACS and BELLS are $\langle z_{defl} \rangle = 0.18$ and 0.50 respectively (Bolton et al. 2008; Brownstein et al. 2012). For CASSOWARY and SL2S, which both used imaging to identify gravitational lenses (Sonnenfeld et al. 2013; Stark et al. 2013), the average deflector redshifts of $\langle z_{spec} \rangle = 0.42$ and 0.49 are also lower than in AGEL (see Table 4).

With our combination of optical and near-infrared spectroscopy, we also confirm background sources with a higher redshift range than existing surveys. SLACS, BELLS, and CASSOWARY confirmed sources with average redshifts up to $z_{src} = 1.2$, 1.5, and 1.76 respectively. For comparison, the average redshift for the AGEL sources is $\langle z_{src} \rangle = 1.91 \pm 0.47$ with confirmed sources up to $z_{src} = 3.549$ (see Figures 7 and 9; Table 4). The lensed sources in AGEL are identified by imaging and complement searches for $z_{src} \geqslant 2$ galaxies based on fiber spectroscopy (e.g., Shu et al. 2016).

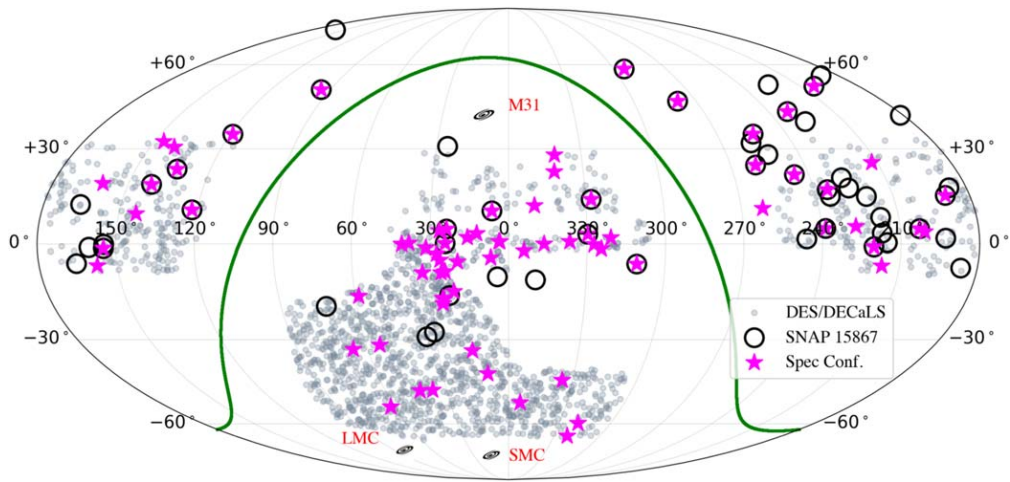

**Figure 3.** Spatial distribution of candidate gravitational lenses in the DES/DECaLS fields (gray circles) and the 77 spectroscopic redshifts from our AGEL survey (pink stars; Table 3) where the secured redshift is of the deflector (foreground) and/or the source (background). The confirmed gravitational lenses span a range in R.A., and most are at declinations near the equator and can be observed by telescopes in both hemispheres; the plane of the Milky Way is shown as the green curve. Several of the confirmed strong lenses are targeted in the HST SNAP program #15867 (open black circles) that provides the high-angular-resolution imaging needed to model the gravitational lenses; additional HST imaging of AGEL systems is ongoing in Cycle 29 (#16773).

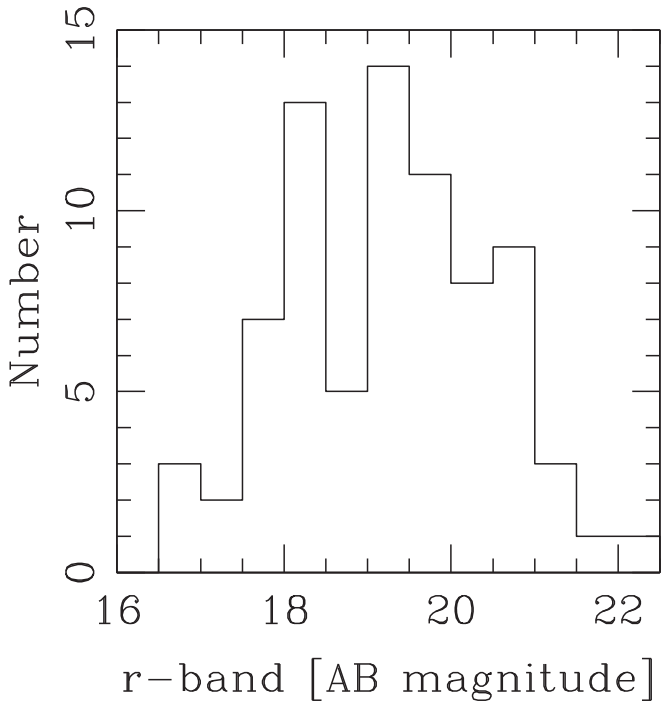

**Figure 4.** Most of the AGEL systems targeted for spectroscopic follow-up are brighter than $r = 21$ mag (foreground deflector; see Figure 1). The total $r$-band magnitudes (AB system) are from the DESI Legacy Survey Data Release 9 and determined using TRACTOR to model the photometry (Dey et al. 2019). For AGEL, we focus mainly on candidate lenses from the DES and DECaLS fields but also include candidates lenses from existing surveys such as CASSOWARY (Stark et al. 2013).

The AGEL survey is a useful resource for recent and ongoing searches that identify thousands of gravitational lens candidates and confirm a subset using spectroscopy. Imaging with the Hyper Suprime-Cam on Subaru has provided an especially rich data set with the SuGOHI team publishing a series of papers identifying a total of ~100 confirmed gravitational lenses and ~1500 possible/probable lenses (Sonnenfeld et al. 2018, 2020; Wong et al. 2018; Jaelani et al. 2020). With the SILO survey, Talbot et al. (2021) identify ~1500 lensing candidates that have BOSS redshifts for the candidate deflectors. With spectroscopic redshifts for deflectors and sources that span the range in redshift, the AGEL survey can be used to estimate contamination in these complementary searches.

Because the J19ab lensing candidates are not limited by fiber diameter and the Einstein radius is proportional to the halo velocity dispersion for an isothermal sphere, we capture a wide range of halo masses including galaxy groups and clusters (see also Huang et al. 2020). Among the confirmed lenses we have seven systems with $r_{EIN} \sim 2''$–$8''$ at $z_{defl} = 0.36$–0.78 (see Figure 1). For comparison, Newman et al. (2015) study 10 strong lensing galaxy groups with $r_{EIN} = 2''.5$–$5''.1$ at $z_{defl} = 0.21$–0.45. Thus our sample extends studies of galaxy groups identified directly by their halo masses ($M_{200} \sim 10^{14}\ M_\odot$) to higher redshifts for comparison to, e.g., cosmological simulations (McCarthy et al. 2017).

### 4.2. Future Science with AGEL Systems

With the AGEL survey, we will provide a rich legacy data set of ~100 strong gravitational lensing systems that can be observed with telescopes in both hemispheres and throughout the year. Such a sample of high-magnification lens systems such as the 68 confirmed in this analysis is ideal for a number of scientific investigations. The data already in hand are being used to study the foreground deflectors and background sources. The ground-based spectroscopy used to confirm the gravitational lenses provides emission-line diagnostics of magnified sources at a key epoch in galaxy formation ($1 < z < 3$; Madau & Dickinson 2014). The width and shape of the spectral lines trace the source kinematics to measure rotation versus dispersion-dominated systems (Leethochawalit et al. 2016; Yuan et al. 2017; Girard et al. 2018; Newman et al. 2018) and search for galactic winds (Jones et al. 2018, Vasan et al. 2022, in preparation). Line ratios

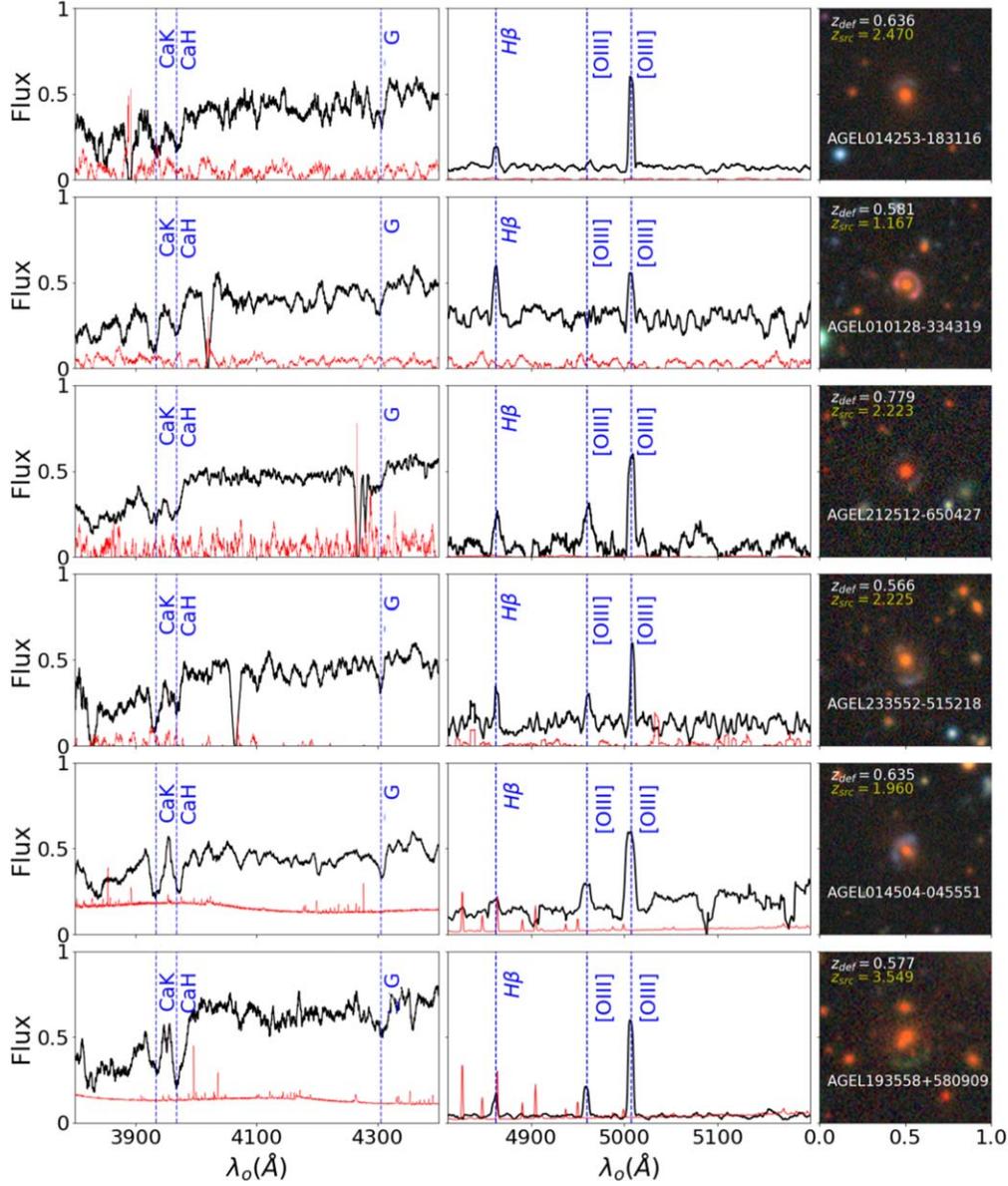

**Figure 5.** By combining optical and near-infrared spectroscopy, we confirm candidate lenses by measuring redshifts for the foreground deflectors and/or background sources. Here are six examples of confirmed gravitational lenses with spectroscopic redshifts for the deflectors (left) and higher-redshift sources (middle); all have redshift quality flag of $Q_z = 3$. The high signal-to-noise ratio (black/red) spectra show strong absorption features for the deflector and emission lines for the source. The RGB images ($26'' \times 26''$; right) are generated from multiband optical imaging from DECam.

Table 4
Comparison to Existing Gravitational Lensing Surveys with Spectroscopic Redshifts

| Survey | Reference | $N_{lenses}$ | $z_{deflector}$ | $\langle z_{deflector} \rangle$ | $z_{source}$ | $\langle z_{source} \rangle$ |
|---|---|---|---|---|---|---|
| AGEL | This paper | 68[a] | 0.21–0.89 | $0.58 \pm 0.14$ | 0.88–3.55 | $1.92 \pm 0.59$ |
| AGEL | This paper | 53[a] | 0.26–0.89 | $0.55 \pm 0.15$ | 0.88–3.55 | $1.91 \pm 0.47$ |
| SLACS | Bolton et al. (2008) | 131 | 0.03–0.51 | 0.18 | 0.09–1.19 | 0.56 |
| BELLS | Brownstein et al. (2012) | 45 | 0.35–0.66 | 0.50 | 0.87–1.52 | 1.17 |
| CASSOWARY | Stark et al. (2013) | 29 | 0.21–0.68 | 0.42 | 0.91–2.81 | 1.76 |
| SL2S | Sonnenfeld et al. (2013) | 35 | 0.23–0.78 | 0.49 | 0.99–3.48 | 2.13 |

**Note.**
[a] We provide statistics for all 68 strong lenses with a secure redshift for either the deflector or source, and for the subset of 53 with secure redshifts for both the deflector and source.

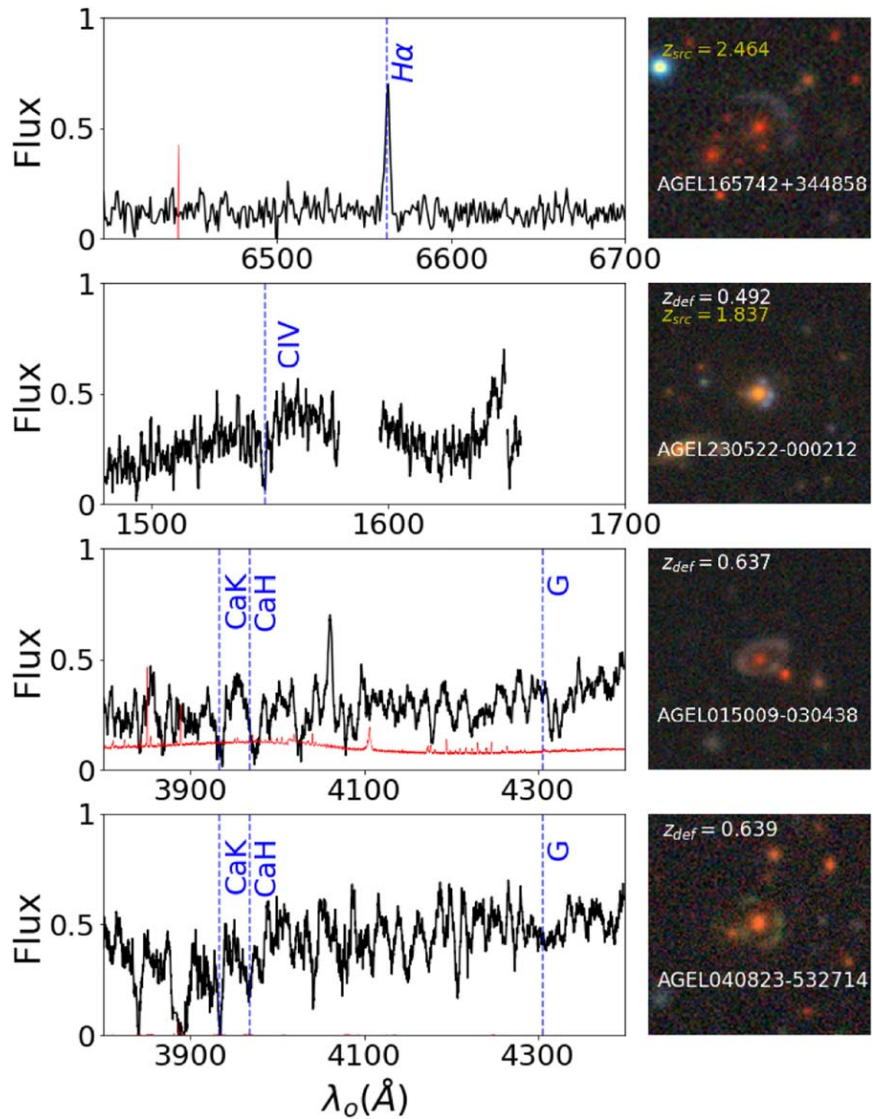

**Figure 6.** Example of spectroscopic redshifts with redshift quality flag $Q_z < 3$. From top to bottom: probable H$\alpha$ emission for source with $Q_z = 1$ (single line); probable C IV absorption with $Q_z = 2.5$ (also weak C III); probable calcium absorption with $Q_z = 1$ (lines not centered); probable calcium absorption with $Q_z = 2$. In our analysis, we use spectroscopic redshifts with $Q_z \geqslant 2$.

such as [N II]/H$\alpha$ and [O III]/H$\beta$ measure gas-phase metallicities and ionization conditions as well as star formation rates and dust content at $z \gtrsim 2$ (e.g., Jones et al. 2012; Sanders et al. 2015, 2016; Tran et al. 2015; Alcorn et al. 2019; Kewley et al. 2019; Harshan et al. 2020)

Gravitational lensing by single galaxies, especially at $z > 0.5$, is particularly effective at testing galaxy formation models. Current cosmological simulations predict that the slope of the mass density profile ($\gamma'$) is essentially flat at $0 < z \lesssim 0.5$ and steepens at $z > 0.5$, but observations of gravitational lenses suggest the opposite is true (Sonnenfeld et al. 2013; Dye et al. 2014). However, most galaxy-scale measurements are at $z < 0.5$, which means the 41 confirmed lenses with deflectors at $z_{\rm defl} \geqslant 0.5$ to date in AGEL provide a key test by increasing the number of systems at $z_{\rm defl} \geqslant 0.5$ (Sonnenfeld et al. 2013).

Increasing the number of confirmed gravitational lenses also enables an exciting range of discovery space such as compound lenses for measuring the Hubble constant and time-variable phenomena for repeated observations via time delays (Suyu et al. 2013, 2017; Kelly et al. 2015). The arcs provide multiple sightlines to probe tomographically the circumgalactic medium of the intervening galaxies (Lopez et al. 2018; Mortensen et al. 2021), including the foreground deflectors. The ∼100 pc-scale measurements that are possible with diffraction-limited observations of lensed sources are particularly relevant for the current and next generation of adaptive optics instruments (Wizinowich et al. 2020) as well as with the James Webb Space Telescope for extending galaxy scaling relations to even lower masses at $z_{\rm src} \geqslant 2$.

### 4.3. High-resolution Imaging with the Hubble Space Telescope

High-resolution imaging is critical for constructing lens models that precisely map the matter distribution of the foreground deflectors. To measure the matter density profiles of the gravitational lenses, we are acquiring high-resolution imaging with the Hubble Space Telescope (#16773; Cycle 29; led by K. Glazebrook) that builds on the existing HST imaging from SNAP program #15867 (Cycle 27; led by X. Huang). By combining the HST imaging with the spectroscopic

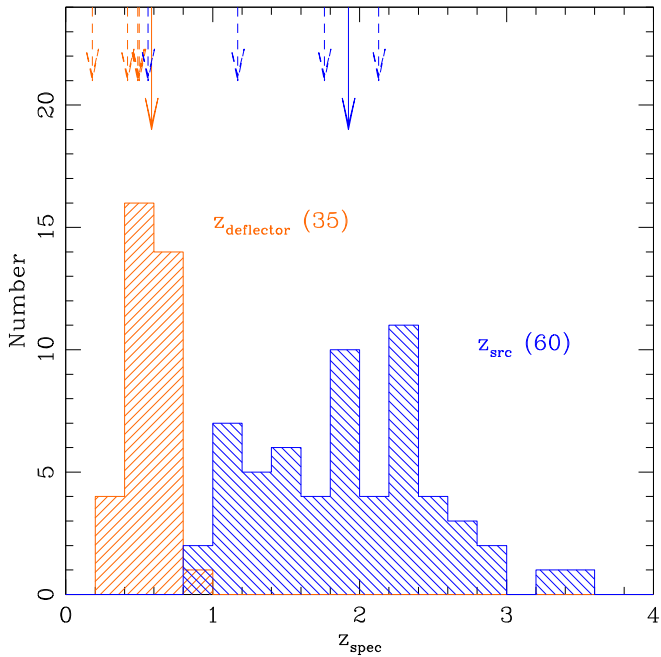

Figure 7. Distribution of our spectroscopic redshifts measured from follow-up with Keck and the VLT; here we show only our measured $z_{spec}$ and exclude the literature redshifts. The deflectors (orange) and sources (blue) have higher average redshifts (solid arrows) relative to SLACS, BELLS, and CASSOWARY (dashed arrows, see Table 4; Bolton et al. 2008; Brownstein et al. 2012; Stark et al. 2013). The SL2S survey based on CFHT imaging (Sonnenfeld et al. 2013) has a higher average source redshift than AGEL, but the spectroscopic ranges for both the deflectors and sources are marginally lower (see Table 4). By combining our spectroscopic redshifts with literature redshifts, we secure redshifts for both $z_{defl}$ and $z_{src}$ for 53 gravitational lenses, and either $z_{defl}$ or $z_{src}$ for 15 lenses.

redshifts measured by AGEL, we will map dark matter substructure and lensed source morphology.

For the HST #16773 observations scheduled through 2023, we target lens candidates at decl. $\lesssim +25°$ to enable follow-up observations by both northern and southern telescopes and candidates that are distributed in R.A. to allow access throughout the year (Figure 3). Targets with existing or scheduled (through 2022) spectroscopic observations for sources and lenses are promoted to higher priority. We also prioritized lens that have existing imaging with the PISCO instrument on Magellan. With the HST observations from the approved programs, we expect to have upwards of 50+ gravitational lenses with HST imaging and spectroscopic redshift for both deflectors and sources by the end of 2023.

## 5. Conclusions

We introduce the ASTRO 3D Galaxy Evolution with Lenses (AGEL) survey by presenting spectroscopically confirmed strong gravitational lenses in the DES and DECaLS fields that are brighter than $r = 22$ mag. In this paper, we report on 79 candidate gravitational lenses selected from a magnitude-limited catalog that were identified in imaging taken with DECam (Figures 1, 3, 4, 5, 6; Jacobs et al. 2019a, 2019b). The combination of deep, high-quality imaging and a search method using convolutional neural networks with human inspection is highly effective at identifying strong lensing systems within the large cosmic volume surveyed by DECam.

We targeted 106 objects for optical–NIR spectroscopy and obtained redshifts for 104 (spectroscopic success rate of 98%).

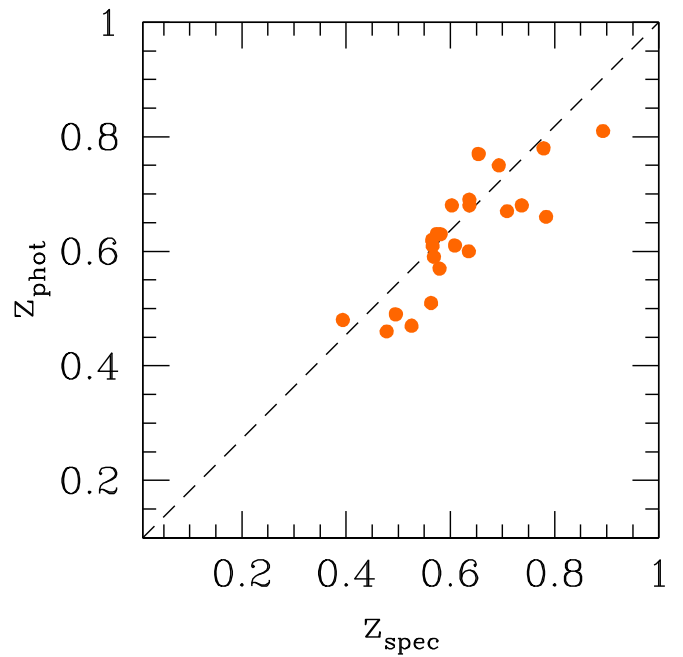

Figure 8. We compare spectroscopic redshifts for the deflectors to photometric redshifts for 23 systems and find that $z_{phot}$ and $z_{spec}$ are remarkably consistent. The average photometric and spectroscopic redshifts are $\langle z_{phot} \rangle = 0.63 \pm 0.10$ and $\langle z_{spec} \rangle = 0.62 \pm 0.11$, and the average absolute difference is $\langle |\delta z|/(1 + z_{spec}) \rangle = 0.03 \pm 0.02$. The photometric redshifts were determined using BPZ and have rms errors of $\Delta z \sim 0.06(1 + z_{spec})$ (Benítez 2000).

Combining our observations with spectroscopic redshifts published in the literature, we have redshifts for 77 candidate lensing systems (Table 3). For 53 lenses, we secure spectroscopic redshifts for both the deflector and source where $z_{src} > z_{defl}$. For 15 lenses, we additionally have eight with $z_{defl} = 0.21–0.79$ and seven with $z_{src} = 1.34–3.39$. Of the remaining nine systems, we identify four as non-lenses while five have inconclusive redshift quality. We define the success rate of the CNN-selected candidate lenses as the ratio 68/77, which is 88%.

The AGEL survey pushes to higher redshifts than previous lensing surveys, with deflectors reaching $z_{defl} \sim 0.9$ and sources spanning a broad redshift range (Figures 5, 7, 9). For the 68 confirmed AGEL systems, the redshift ranges for the foreground deflectors and background sources are $z_{defl} = 0.21–0.89$ and $z_{src} = 0.88–3.55$, and the average redshifts are $\langle z_{defl} \rangle = 0.58 \pm 0.14$ and $\langle z_{src} \rangle = 1.92 \pm 0.59$. There are 41 strong lenses with deflectors at $z_{defl} \geq 0.5$. The resulting sample is well suited for addressing a range of questions in astrophysics and cosmology such as the current uncertainty of whether mass density profiles evolve with redshift.

The AGEL survey provides a useful training set to further refine automated all-sky searches for strong gravitational lenses, especially given the high purity of the CNN-selected sample. For the subset of 23 confirmed lenses with photometric redshifts from existing surveys (Figure 8), the photometric redshifts are remarkably consistent with the spectroscopic redshift of the deflector: the average absolute difference is $\langle |\delta z|/(1 + z_{spec}) \rangle = 0.03 \pm 0.02$. However, spectroscopy of the candidate deflectors and sources remains critical to confirming whether the system is a strong gravitational lens.

Our goal is to spectroscopically confirm a statistically robust sample of $\sim 100$ strong gravitational lenses that can be observed with adaptive optics using telescopes in both hemispheres throughout the year. The optical/NIR spectroscopy combined with existing multiwavelength observations in the DES and DECaLS fields already enables a wide range of studies such as measuring the total matter profiles of the foreground deflectors, using multiple sightlines to probe the circumgalactic medium, and searching for galactic-scale winds in the background sources. In order to more accurately model the lens mass distribution, spatially resolve subkiloparsec structure in the sources, and search for dark matter substructure in the deflectors and along the line of sight, we are also acquiring high-resolution imaging with the Hubble Space Telescope (#GO-16773) for a subset of AGEL systems.

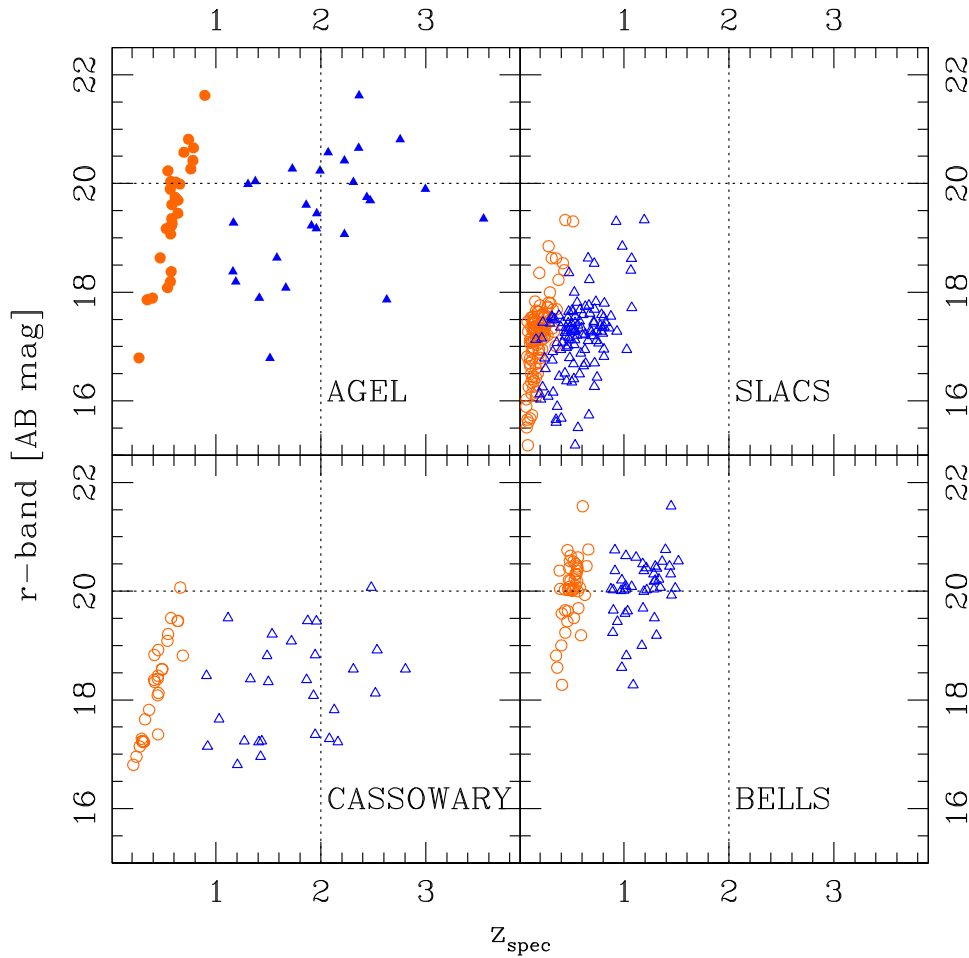

**Figure 9.** We compare the total *r* magnitudes reported in DESI DR9 for AGEL, SLACS, BELLS, and CASSOWARY to the spectroscopic redshifts for deflectors (circles) and sources (triangles). The DESI DR9 photometry is deeper and of higher resolution than SDSS and is thus able to deblend systems reported in earlier surveys. The lenses from the earlier surveys (open symbols) tend to be brighter than those in AGEL; only BELLS reaches comparable depths. The combination of deeper imaging with higher angular resolution and optical–NIR spectroscopy results in AGEL deflectors (orange filled circles) and sources (blue filled triangles) that reach higher redshifts.

We thank the referee for a detailed and constructive report. Data presented herein were obtained at the W. M. Keck Observatory, which is operated as a scientific partnership among the California Institute of Technology, the University of California and the National Aeronautics and Space Administration. The authors wish to recognize and acknowledge the very significant cultural role and reverence that the summit of Maunakea has always had within the indigenous Hawaiian community. We are most fortunate to have the opportunity to conduct observations from this mountain. Data include observations collected at the European Organisation for Astronomical Research in the Southern Hemisphere under ESO program 0101.A-0577. The authors acknowledge support by the Australian Research Council Centre of Excellence for All-Sky Astrophysics in 3 Dimensions (ASTRO 3D), through project number CE170100013. S.L. is funded by FONDECYT grant number 1191232. T.J. and K.V.G.C. gratefully acknowledge funding support for this work from the Gordon and Betty Moore Foundation through Grant GBMF8549, and from a Dean's Faculty Fellowship. T.J. acknowledges support from the National Science Foundation through grant AST-2108515. T.E.C. is funded by a Royal Society University Research Fellowship and from the European Research Council (ERC) under the European Union's Horizon 2020 research and innovation program (LensEra: grant agreement No 945536).

*Facilities:* W. M. Keck Observatory, European Southern Observatory, Cerro Tololo Inter-American Observatory

**ORCID iDs**

Kim-Vy H. Tran ⓘ https://orcid.org/0000-0001-9208-2143
Anishya Harshan ⓘ https://orcid.org/0000-0001-9414-6382
Karl Glazebrook ⓘ https://orcid.org/0000-0002-3254-9044
Tucker Jones ⓘ https://orcid.org/0000-0001-5860-3419
Colin Jacobs ⓘ https://orcid.org/0000-0003-4239-4055

Glenn G. Kacprzak https://orcid.org/0000-0003-1362-9302
Tania M. Barone https://orcid.org/0000-0002-2784-564X
Thomas E. Collett https://orcid.org/0000-0001-5564-3140
Anshu Gupta https://orcid.org/0000-0002-8984-3666
Lisa J. Kewley https://orcid.org/0000-0001-8152-3943
Sebastian Lopez https://orcid.org/0000-0003-0389-0902
Themiya Nanayakkara https://orcid.org/0000-0003-2804-0648
Ryan L. Sanders https://orcid.org/0000-0003-4792-9119
Sarah M. Sweet https://orcid.org/0000-0002-1576-2505